\documentclass[a4paper]{jpconf}
\usepackage{graphicx}
\usepackage{bm}
\usepackage{braket}

\usepackage{color}

\begin{document}
\title{Toroidal orders and related phenomena \\ in nonmagnetic and magnetic materials}

\author{Hiroaki Kusunose$^{1,2}$ \& Satoru Hayami$^3$}

\address{$^1$Department of Physics, Meiji University, Kawasaki 214-8571, Japan}
\address{$^2$Quantum Research Center for Chirality, Institute for Molecular Science, Okazaki 444-8585, Japan}
\address{$^3$Graduate School of Science, Hokkaido University, Sapporo 060-0810, Japan}
\ead{hk@meiji.ac.jp}

\begin{abstract}
In this short article, we overview a concept of electronic toroidal multipoles, and their ordering with associated physical properties in non-magnetic and magnetic materials.
The toroidal multipoles are introduced as microscopic electronic variables in view of symmetry.
They are classified according to crystallographic and magnetic point groups, which allows us to discuss various possible cross correlations in a transparent and unified manner.
The representative examples of toroidal orders and related phenomena, and the mutual relationship between these orders are given, with focusing on monopoles and dipoles.
The concept of toroidal multipoles would promote future studies toward observations and identifications of unknown electronic phases and their related physical phenomena.
\end{abstract}

\section{Introduction}

Symmetry of materials and their physical properties are closely related.
Representative examples that spontaneously break fundamental symmetries, i.e., the time-reversal and spatial inversion symmetries, are ferromagnetism and ferroelectricity, respectively.
They are long-range order of microscopic variables, magnetic (M) and electric (E) dipoles.
In recent years, alternative orders have been extensively investigated in which both time-reversal and spatial inversion symmetries are broken~\cite{van2007observation, Spaldin_0953-8984-20-43-434203, kopaev2009toroidal} or both of them are not broken~\cite{jin2020observation, hayashida2020visualization}.
They are called as toroidal or ferroaxial orders, respectively.
These orders are characterized by the alignment of a certain type of microscopic variables as well, which are the main subject of the present paper.

In contrast to an electron in vacuum, electrons in materials have a rich variety of degrees of freedom, i.e., charge, orbital, spin, and sublattice in underlying molecular or crystal structures.
By considering their combinations, various microscopic variables other than ordinary electric and magnetic dipoles can be constructed, which become order parameters of unconventional orders and bring about associated exotic physical properties.

In the classification of electronic degrees of freedom in materials, two types of multipoles, E and M multipoles have been recognized~\cite{Kusunose_JPSJ.77.064710, kuramoto2009multipole, Santini_RevModPhys.81.807}.
Subsequently, additional two types of multipoles, the electric toroidal (ET) and magnetic toroidal (MT) multipoles, that are indispensable for a unified description of a state of matter, have been incorporated~\cite{dubovik1986axial, dubovik1990toroid, hayami2018microscopic, Hayami_PhysRevB.98.165110, kusunose2020complete, hayami2024unified}.
The ET and MT multipoles have the opposite spatial-inversion parity to the conventional E and M multipoles, and these four types of multipoles (E, M, ET, and MT) constitute a complete basis set to describe arbitrary electronic degrees of freedom in materials in molecules and crystals~\cite{Kusunose_PhysRevB.107.195118}.
For example, ET and MT dipoles are the microscopic variables of ferroaxial and toroidal orders, respectively~\cite{Spaldin_0953-8984-20-43-434203, kopaev2009toroidal, Hayami_doi:10.7566/JPSJ.91.113702}.
On the other hand, the ET monopole having time-reversal even pseudoscalar property is a practical entity of chirality~\cite{Hayami_PhysRevB.98.165110, kusunose2020complete, Oiwa_PhysRevLett.129.116401, kishine2022definition,Kusunose2024}, which was thoughtfully discussed by Barron~\cite{barron1986symmetry, barron1986true, Barron_2004}.
Moreover, the MT monopole can be realized in certain types of antiferromagnet (AFM)~\cite{Hayami_PhysRevB.108.L140409}, which will be discussed in this paper.

This paper is organized as follows.
In Sec.~\ref{sec2}, we introduce the concept of toroidal multipole from symmetry viewpoint at the classical and quantum-mechanical level.
Then, we discuss the toroidal orders and related phenomena, with focusing on monopoles and dipoles in Sec.~\ref{sec3}.
The relation between point groups and active multipoles, and possible cross correlations under multipole orders are also clarified.
In Sec.~\ref{sec4}, we give prime examples for ET and MT monopole and dipole orders and related phenomena.
The final section summarizes the paper.

\section{Concept of toroidal multipole}\label{sec2}

\subsection{Multipole and symmetry}

\begin{table}[t!]
\caption{
\label{tbl1}
Symmetry properties of the four types of monopoles and dipoles.
$\mathcal{M}_{\perp}$ ($\mathcal{M}_{\parallel}$) represents the mirror operation with respect to the plane perpendicular (parallel) to dipole.
$\bar{\bm{r}}=\bm{r}/|\bm{r}|$, $\bm{l}$, and $\bm{\sigma}$ represent dimensionless position vector, orbital angular momentum, and Pauli matrix, respectively.
The definitions of operators, $\bm{Q}'$, $\bm{G}'$, and $\bm{T}'$, are given by a product of quadrupoles and spins, and their expressions are complicated as given in \cite{kusunose2020complete}.
\#PG and \#MG are the number of point and magnetic point groups allowing the corresponding multipole orders.
}
\scriptsize
\begin{center}
\begin{tabular}{ccccccccccc}
\br
MP & $\mathcal{R}$ & $\mathcal{T}$ & $\mathcal{P}$ & $\mathcal{M}_{\perp}$ & $\mathcal{M}_{\parallel}$ & \#PG & \#MPG & Form (orbital) & \multicolumn{2}{c}{Form (spin)} \\
\br
$Q_{0}$ & $l=0$ & $+$ & $+$ & $+$ & $+$ & $32$ & $122$ & $1$ & $\frac{1}{\sqrt{3}}(\bm{l}\cdot\bm{\sigma})$ \\
$M_{0}$ & $l=0$ & $-$ & $-$ & $-$ & $-$ &  & $32$ & --- & $\frac{1}{\sqrt{3}}(\bar{\bm{r}}\cdot\bm{\sigma})$ \\
$G_{0}$ & $l=0$ & $+$ & $-$ & $-$ & $-$ & $11$ & $32$ & --- & $\frac{1}{\sqrt{3}}(\bm{t}\cdot\bm{\sigma})$ \\
$T_{0}$ & $l=0$ & $-$ & $+$ & $+$ & $+$ & & $32$ & $i$ (off-diagonal) & --- \\
\mr
$\bm{Q}$ & $l=1$ & $+$ & $-$ & $-$ & $+$ & $10$ & $31$ & $\bar{\bm{r}}$ & $\frac{1}{\sqrt{3}}(\bm{\sigma}\times\bm{t})$ & $\bm{Q}'$ \\
$\bm{M}$ & $l=1$ & $-$ & $+$ & $+$ & $-$ & & $31$ & $\bm{l}$ & $\bm{\sigma}$ & $\frac{1}{\sqrt{10}}[3(\bar{\bm{r}}\cdot\bm{\sigma})\bar{\bm{r}}-\bar{\bm{r}}^{2}\bm{\sigma}]$ \\
$\bm{G}$ & $l=1$ & $+$ & $+$ & $+$ & $-$ & $13$ & $43$ & --- & $\frac{1}{\sqrt{2}}(\bm{\sigma}\times\bm{l})$ & $\bm{G}'$ \\
$\bm{T}$ & $l=1$ & $-$ & $-$ & $-$ & $+$ & & $31$ & $\bm{t}\equiv\frac{1}{6}(\bar{\bm{r}}\times\bm{l})+{\rm h.c.}$ & $\frac{1}{\sqrt{2}}(\bm{\sigma}\times\bar{\bm{r}})$ & $\bm{T}'$ \\
\br
\end{tabular}
\end{center}
\end{table}

It is well known that a charge distribution brings about a polar quantity such as an electric dipole $\bm{Q}$, while an electric current distribution yields an axial quantity such as a magnetic dipole $\bm{M}$.
Once such dipoles exist, a vortex-like alignment of them gives us another independent quantity characterized by a dipole ($\bm{G}\propto\sum_{j}\bm{R}_{j}\times\bm{Q}_{j}$ or $\bm{T}\propto\sum_{j}\bm{R}_{j}\times\bm{M}_{j}$ where $\bm{R}_{j}$ is a position of the dipole), which has an opposite spatial parity of $\bm{Q}$ or $\bm{M}$ as the cross product contains the totally anti-symmetric (Levi-Civita) tensor.
This is the so-called electric-toroidal (ET) or magnetic-toroidal (MT) dipole, representing the degree of vorticity of a distribution of electric or magnetic dipoles.
The concept of toroidal dipoles (rank $l=1$ with the component $m$) can be generalized straightforwardly to any ranks (called a toroidal multipole $G_{lm}$ or $T_{lm}$) introduced as a classical quantity in the middle of 1980s~\cite{dubovik1986axial, dubovik1990toroid}.
In this paper, we use the term ``multipole'' to indicate a whole set of four types of multipoles, i.e., $Q_{lm}$, $M_{lm}$, $T_{lm}$, and $G_{lm}$, or for the conventional $Q_{lm}$ and $M_{lm}$ in a specific context, while we always use the term ``toroidal multipole'' to indicate $T_{lm}$ or $G_{lm}$.

The concept of multipole is closely related to the symmetry.
The principal symmetry of interest for materials science consists of the time-reversal $\mathcal{T}$, spatial inversion $\mathcal{P}$, and rotation $\mathcal{R}$, while the translational operation is less important in the present argument.
As similar to the multipole expansion for characterizing an anisotropic distribution of electromagnetic fields, an anisotropic state of materials can be characterized by ``multipoles'' with a set of properties of $(\mathcal{T},\mathcal{P},\mathcal{R})$, i.e., ``electric$(+)$/magnetic$(-)$'' of $\mathcal{T}$, ``polar[$(-1)^{l}$]/axial[$(-1)^{l+1}$]'' of $\mathcal{P}$, and $(l,m)$ of $\mathcal{R}$.
The enumeration of all combinations corresponds either to $Q_{lm}$ (electric polar), $M_{lm}$ (magnetic axial), $G_{lm}$ (electric axial), or $T_{lm}$ (magnetic polar), and hence, these four-type of multipoles constitute a symmetry-adapted complete basis set~\cite{hayami2018microscopic, Hayami_PhysRevB.98.165110, kusunose2020complete, hayami2024unified, Kusunose_PhysRevB.107.195118}.
In other words, a state of materials is described by a linear combination of these four-type multipole bases as a building block with an emphasis on the symmetry.
For instance, the properties of the multipoles for $l=0,1$ are summarized in Table~\ref{tbl1}.
It is noted that the mirror operation $\mathcal{M}(\bm{n})$ with respect to a plane perpendicular to a unit vector $\bm{n}$ is equivalent to $\mathcal{P}$ in addition to $180^{\circ}$ rotation around $\bm{n}$ [$C_{2}(\bm{n})$], i.e., $\mathcal{M}(\bm{n})=\mathcal{P}C_{2}(\bm{n})$.
Therefore, $\mathcal{P}$ and $\mathcal{M}(\bm{n})$ have no differences for monopole, while $\mathcal{P}$ and $\mathcal{M}(\bm{n})$ for dipole with $\bm{n}$ being perpendicular to it ($\mathcal{M}_{\parallel}$) differ since $C_{2}(\bm{n})$ gives an additional sign for dipole.

\subsection{Symmetry lowering and emergence of multipole}

As long as a system is invariant under $(\mathcal{T},\mathcal{P},\mathcal{R})$ operations, an electric monopole [$l=m=0$, $\mathcal{T}=+$, $(\mathcal{P},\mathcal{M}_{\perp},\mathcal{M}_{\parallel})=(+,+,+)$], i.e., electric charge, is only allowed to exist.
However, since some of $(\mathcal{T},\mathcal{P},\mathcal{R})$ is lost in realistic materials, the corresponding multipoles other than $Q_{0}$ can appear.
In other words, the appearance of multipoles characterizes how the symmetry of a system is lost.
In the case of macroscopic systems, such symmetry-adapted multipole represents a possible candidate of order parameter, while for an isolated molecule, it characterizes its wave function in terms of symmetry.
Note that $\mathcal{P}$ or $\mathcal{M}(\bm{n})$ alone can exist in the case without $C_{2}(\bm{n})$ in some point groups, in contrast to the rotation group where $C_{2}(\bm{n})$ and hence both $\mathcal{P}$ and $\mathcal{M}(\bm{n})$ always exist.

\begin{figure}[tb]
\centering
\includegraphics[width=14cm]{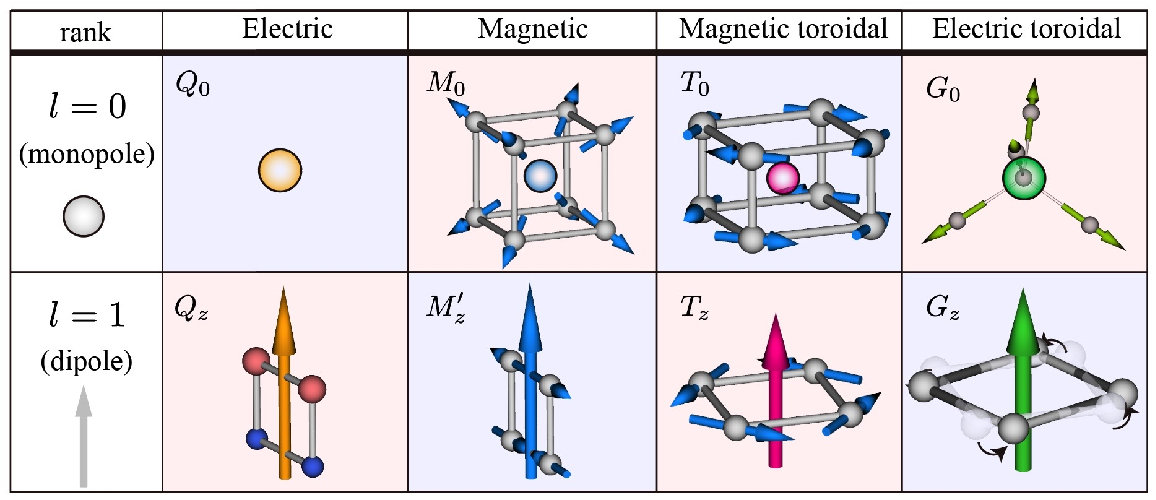}
\vspace{3mm}
\caption{\label{fig:cluster}
Four types of cluster multipoles up to rank 1.
The orange, blue, red, and green objects represent electric (E), magnetic (M), magnetic-toroidal (MT), and electric-toroidal (ET) multipoles, respectively.
The cluster consisting of primitive objects at sublattice sites represents a cluster multipole. For example, the static structural rotational distortion (E dipole) in the lower most-right panel corresponds to a cluster ET dipole. The blue and red backgrounds indicate even- and odd-parity quantities, respectively.
}
\end{figure}

As shown above, the toroidal multipole was introduced as a classical quantity, where the dipole moment, $\bm{Q}_{j}$ in $\bm{G}$ or $\bm{M}_{j}$ in $\bm{T}$, is the expectation value of the corresponding quantum-mechanical electric or magnetic dipole operator.
Such a classical concept can straightforwardly be connected to the cluster-type multipoles in sublattice systems as shown in Fig.~\ref{fig:cluster}.
Therefore, a certain type of antiferromagnets accompany the ferroic ordering of $\bm{T}$ (e.g., $T_{z}$ as shown in Fig.~\ref{fig:cluster}) which breaks $(\mathcal{T},\mathcal{P},\mathcal{M}_{\perp})$.
Such antiferromagnets have been investigated extensively, referred to as various terms such as magneto-electric, (magnetic) toroidal, anapole, or MT-dipole ordering.
Precisely speaking, an MT state is characterized by a component of $\bm{T}$ belonging to the identity representation in the ordered phase.
UNi$_{4}$B~\cite{saito2018evidence, ota2022zero, Hayami_PhysRevB.90.024432, Ishitobi_PhysRevB.107.104413}, $\alpha$-Cu$_2$V$_2$O$_7$~\cite{Gitgeatpong_PhysRevB.92.024423,Gitgeatpong_PhysRevB.95.245119,Gitgeatpong_PhysRevLett.119.047201,Hayami_doi:10.7566/JPSJ.85.053705}, and CuMnAs~\cite{wadley2013tetragonal, wadley2016electrical, Wang_PhysRevLett.127.277201} are the prime examples showing the MT-dipole ordering.
Since the MT-dipole ordering has already been investigated extensively, we will focus on the discussions about other toroidal monopole and dipole orderings ($G_{0}$, $T_{0}$, and $\bm{G}$) here, and we only show the literatures~\cite{Spaldin_0953-8984-20-43-434203,kopaev2009toroidal} for the MT-dipole ($\bm{T}$) orderings.

Meanwhile, since the MT monopole $T_{0}$ has properties $(\mathcal{P},\mathcal{M}_{\perp},\mathcal{M}_{\parallel})=(+,+,+)$ and $\mathcal{T}=-$, an emergence of $T_{0}$ breaks only $\mathcal{T}$ symmetry, which mixes between time-reversal partners, i.e., $Q_{lm}$ and $T_{lm}$ or $G_{lm}$ and $M_{lm}$ without changing other symmetry properties.
This time-reversal switching aspect due to $T_{0}$ leads to intriguing cross correlations as shown later~\cite{Hayami_PhysRevB.108.L140409}.
From a symmetry point of view, $T_{0}$ is expressed as $T_{0}\propto \sum_{j}\bm{R}_{j}\cdot\bm{T}_{j}$ for example, which indicates that a vortex-like magnetic structure has a divergent property (e.g., $T_{0}$ as shown in Fig.~\ref{fig:cluster}).
The practical examples will be discussed later.

Similarly, a vortex-like alignment of electric dipole moments accompanies a ferroic ordering of $\bm{G}$.
Since it corresponds to a static structural rotational distortion ($G_{z}$ as shown in Fig.~\ref{fig:cluster}), it is called as rotational, ferroaxial, or ET-dipole ordering.
The gyrotropic vector, which plays a significant role in optical phenomena in anisotropic media, belongs to the same category of $\bm{G}$ from the symmetry of view.
In this case, a component of $\bm{G}$ becomes the identity representation in the symmetry-lowering state, which breaks the vertical mirror $\mathcal{M}_{\parallel}$ with keeping $\mathcal{T}$, $\mathcal{P}$, and $\mathcal{M}_{\perp}$ as $\bm{G}$ has the properties $(\mathcal{T},\mathcal{P}, \mathcal{M}_{\perp})=(+,+,+)$ and $\mathcal{M}_{\parallel}=-$.
RbFe(MoO$_4$)$_2$~\cite{jin2020observation, Hayashida_PhysRevMaterials.5.124409}, NiTiO$_3$~\cite{hayashida2020visualization, Hayashida_PhysRevMaterials.5.124409, fang2023ferrorotational}, and Ca$_5$Ir$_3$O$_{12}$~\cite{Hasegawa_doi:10.7566/JPSJ.89.054602, Hanate_doi:10.7566/JPSJ.89.053601, hanate2021first, hanate2023space} are the prime examples showing the ET-dipole ordering.

Moreover, chirality is characterized by pure rotation operations only with keeping the time-reversal symmetry, and there are no mirror and spatial inversion operations~\cite{barron1986true, Barron_2004, kelvin1894molecular, kelvin2010baltimore}.
This type of symmetry breaking corresponds to the emergence of $G_{0}$ which has the properties $\mathcal{T}=+$ and $(\mathcal{P},\mathcal{M}_{\perp},\mathcal{M}_{\parallel})=(-,-,-)$~\cite{Hayami_PhysRevB.98.165110, kusunose2020complete, Oiwa_PhysRevLett.129.116401, kishine2022definition,Kusunose2024}.
As $G_{0}$ is $\mathcal{T}$-even axial monopole (pseudoscalar), one possible realization of chirality is characterized by $G_{0}\propto\sum_{j}\bm{R}_{j}\cdot\bm{G}_{j}$ for example as shown in the upper most-right panel in Fig.~\ref{fig:cluster}, which implies a three-dimensional structure containing a divergent character of a vortex-like alignment of electric dipole moments corresponding to structural chirality.

As shown in the above arguments, a toroidal monopole represents a degree of divergence of corresponding dipole distribution, while a toroidal dipole represents a degree of vorticity (rotation) of opposite-parity dipole distribution.
They are comprehensive information of dipole distribution of materials.

\subsection{Quantum-mechanical operator for toroidal multipole}

\begin{figure}[tb]
\centering
\includegraphics[width=14cm]{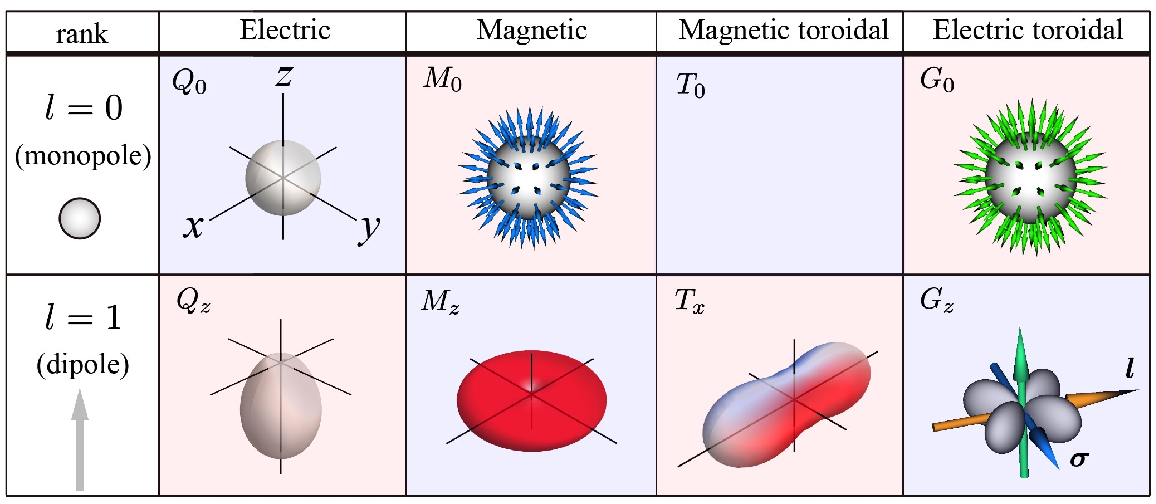}
\vspace{3mm}
\caption{\label{fig:atomic}
Four types of multipoles within a single atom up to rank 1.
The orange and blue arrows represent the orbital $\bm{l}$ and spin $\bm{\sigma}$ angular-momentum operators, respectively, while the orange arrow is the ET dipole operator, $\bm{G}\propto\bm{l}\times\bm{\sigma}$ in the lower most-right panel.
The electric-density distribution of the wave function is indicated by the gray shape, while the red-blue colormap indicates the distribution of $l_{z}$.
The blue and green arrows in the upper panel represent the divergent flux-type distributions of $\bm{\sigma}$ and $\bm{G}$.
The blue and red backgrounds indicate even- and odd-parity quantities, respectively.
}
\end{figure}

In contrast to the above classical or cluster view of toroidal multipoles, the quantum-mechanical operators of them within a single-site atomic wave function were found recently~\cite{hayami2018microscopic, kusunose2020complete}.
The expressions of monopoles and dipoles with or without the spin degrees of freedom are summarized in Table~\ref{tbl1}, and the complete expressions of the higher-rank atomic-scale multipoles in terms of spherical harmonics and their derivatives are systematically found in literature~\cite{hayami2018microscopic, kusunose2020complete, Kusunose_PhysRevB.107.195118}.
The four types of monopoles and dipoles in a single atom are schematically illustrated in Fig.~\ref{fig:atomic}.
Note that $T_{0}$ cannot be defined within a single atomic wave function.
It should be emphasized that the definition of the four types of \textit{cluster} multipoles is obtained systematically by replacing the the position vector $\bm{r}$ in the definition of the \textit{atomic-scale} multipoles in Table~\ref{tbl1} with the position vectors, $\bm{R}_{i}$, in the cluster~\cite{Kusunose_PhysRevB.107.195118,Suzuki_PhysRevB.99.174407,Suzuki_PhysRevB.95.094406}.
Such a replacement works out because the change of length scale does not change the symmetry.

Although the direct observation for the atomic-scale toroidal multipoles as shown in the above has not been succeeded so far, such microscopic degrees of freedom could be essential ingradients in understanding various cross correlations at microscopic quantum-mechanical level~\cite{hayami2018microscopic,kishine2022definition,hayami2024unified,Kusunose2024}.
For clarity of terminology, we use in the following ``atomic'' multipoles for describing electronic degrees of freedom in terms of orbital and/or spin angular momentum, ``site-cluster'' multipoles for composite ones consisting of atomic multipoles at several sites in a cluster unit, and ``bond-cluster'' multipoles for ones on a bond connecting between two sites.

\section{Toroidal ordering and related phenomena}\label{sec3}

In this section, we introduce the relationship between the 32 crystallographic and 122 magnetic point groups and active multipoles with $l=0,1$, and the expected cross correlations in the presence of active multipoles.

\subsection{Crystallographic point groups and multipoles}

The 32 crystallographic point groups (C-PGs) are defined by the group of symmetry operations $\mathcal{G}$ from $\mathcal{R}$, $\mathcal{P}$ and their combinations, i.e., roto-inversion operation, $\mathcal{RP}$ [$\mathcal{G}\in (\mathcal{R},\mathcal{P},\mathcal{RP})$].
Note that the mirror and roto-reflection operations can be expressed by $\mathcal{RP}$.
Since $\mathcal{T}$ is not included in the symmetry operations of C-PG, there is no meaning to distinguish between electric and magnetic properties.
Therefore, we discuss the electric multipoles in this subsection by assuming non-magnetic systems.

\begin{figure}[tb]
\centering
\includegraphics[width=15cm]{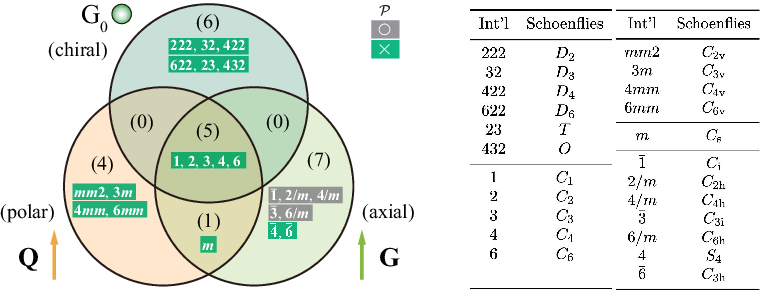}
\caption{\label{fig1}
Relation between crystallographic point groups and active electric-type multipoles.
The background color of each point group represents the presence or absence of $\mathcal{P}$ symmetry.
The additional $\mathcal{P}$-odd multipoles ($l>1$) exist in $\bar{4}$ and $\bar{6}$.
The correspondence between the international and Schoenflies symbols is also shown.
}
\end{figure}

The relation between C-PGs and active E dipole ($\bm{Q}$) and ET monopole ($G_{0}$) and dipole ($\bm{G}$) is shown in Fig.~\ref{fig1} where the number in parenthesis in the Venn diagram indicates that of C-PGs that activates each multipole~\cite{Hayami_PhysRevB.98.165110}.
The total number of C-PGs for each multipole is listed in Table~\ref{tbl1}, where $Q_{0}$ is always active in 32 C-PGs.
Since $\bm{G}$ itself does not break the inversion symmetry, the noncentrosymmetric PGs in the $\bm{G}$ category (indicated by grey background) should accompany higher-rank odd-parity multipoles.
There are 23 C-PGs in Fig.~\ref{fig1}, and the rest of nine C-PGs have active multipoles with $l>1$.

\subsection{Magnetic point groups and multipoles}

The 122 magnetic point groups (M-PGs) are defined by the combination of $\mathcal{T}$ and symmetry operations of C-PG.
They are further classified into three types depending on how $\mathcal{T}$ is included~\cite{bradley2009mathematical}.
Type I (32 groups) is the same as C-PGs where $\mathcal{T}$ is completely lost, which is indicated by no operations with prime.
Type II (32 groups) called as grey group has the symmetry operations $\mathcal{TG}$ in addition to $\mathcal{G}$.
This type of M-PG is expressed by attaching $1'$ to the corresponding C-PG.
The rest of 58 M-PGs are in Type III called as black-and-white (BW) PG.

\begin{figure}[tb]
\centering
\includegraphics[width=15cm]{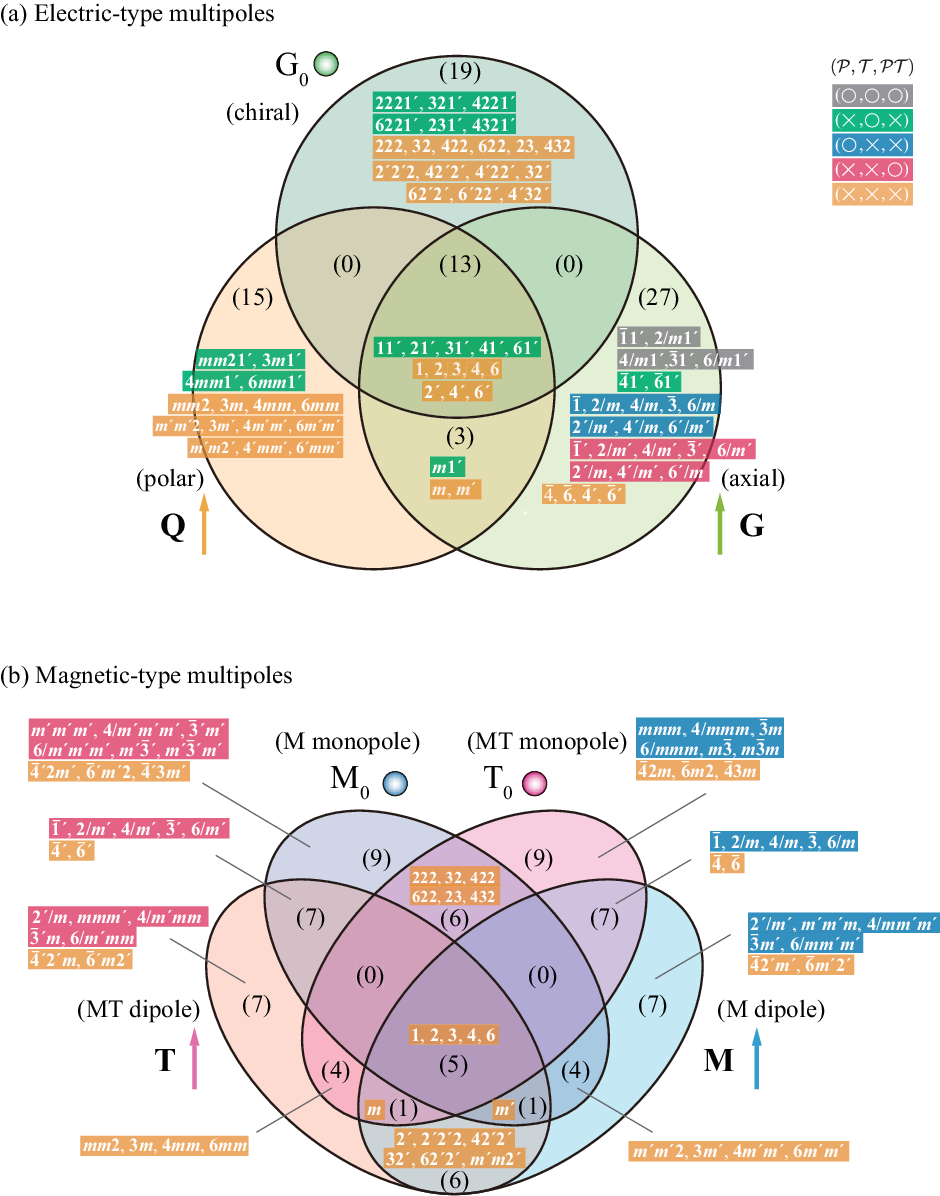}
\caption{\label{fig2}
Relation between magnetic point groups and active multipoles.
(a) the electric-type multipoles and (b) the magnetic-type multipoles.
The background color of each magnetic point group represents the presence or absence of $(\mathcal{P},\mathcal{T},\mathcal{PT})$ symmetries.
The additional $\mathcal{P}$-odd multipoles ($l>1$) exist in the noncentrosymmetric point groups of pure $\bm{G}$, $T_{0}$, and $\bm{M}$ categories.
}
\end{figure}

Since the presence or absence of $\mathcal{T}$ is taken into account in M-PGs, there is a proper meaning of electric-type and magnetic-type multipoles.
The relation between M-PGs and active monopoles and dipoles are shown in Fig.~\ref{fig2} for (a) electric-type multipoles and (b) magnetic-type multipoles, where the number in parenthesis in the Venn diagram indicates that of M-PGs that activates each multipole~\cite{Yatsushiro_PhysRevB.104.054412}.
The presence or absence of $(\mathcal{P},\mathcal{T},\mathcal{PT})$ is indicated by the background color of each M-PG.
There are $77$ and $73$ M-PGs in Fig.~\ref{fig2}(a) and (b), respectively, and the rest of $45$ and $49$ M-PGs have active electric-type and magnetic-type multipoles with $l>1$.

The nonmagnetic states accompanying only electric-type multipoles should belong to the Type II grey groups (grey or green backgrounds) which do not appear in Fig.~\ref{fig2}(b).
For example, 11 chiral states in Type II (green background) are nonmagnetic chiral states, while chiral states such as $222$ in Type I and $2'2'2$ in Type III coexist other magnetic-type multipoles, like $M_{0}$ and $T_{0}$ for $222$ and $\bm{M}$ and $\bm{T}$ for $2'2'2$ as shown in Fig.~\ref{fig2}(b).
Note that there are symmetry correspondences, $T_{0}M_{0}$, $\bm{T}\cdot\bm{M}\leftrightarrow G_{0}$, indicating that $G_{0}$ coexists with $\bm{T}$ and $\bm{M}$ that are parallel to each other in $2'2'2$.
Similarly, 13 axial states in Type II (grey or green backgrounds) are nonmagnetic axial states, while axial states such as $\bar{1}$ in Type I and $\bar{1}'$ in Type III coexist with ($M_{0},\bm{T}$) and ($T_{0},\bm{M}$), respectively, as the symmetry correspondence $\bm{G}\leftrightarrow T_{0}\bm{M}\leftrightarrow M_{0}\bm{T}$ holds.
Moreover, $2'/m$ and $2'/m'$ in Type III coexist $\bm{T}$ and $\bm{M}$, respectively, where the symmetry correspondence $\bm{G}\leftrightarrow \bm{T}\times\bm{T}$ or $\bm{G}\leftrightarrow \bm{M}\times\bm{M}$ indicates two distinct vectors ($\bm{T}_{1},\bm{T}_{2}$) or ($\bm{M}_{1},\bm{M}_{2}$) are orthogonal with each other in $2'/m$ and $2'/m'$, respectively.
On the contrary, there is the symmetry correspondences $\bm{Q}\leftrightarrow M_{0}\bm{M}\leftrightarrow T_{0}\bm{T}$ and $\bm{Q}\leftrightarrow\bm{T}\times\bm{M}$.
Thus, the M-PGs in the intersections of $(M_{0},\bm{M})$ and $(T_{0},\bm{T})$ in Fig.~\ref{fig2}(b) appear in $\bm{Q}$ in Fig.~\ref{fig2}(a), and $\bm{T}$ and $\bm{M}$ are orthogonal to each other in $m'm2'$ for example.

It is noted that since $(M_{0}$,$T_{0}$) or $(\bm{M}$,$\bm{T}$) have opposite properties of ($\mathcal{P},\mathcal{M}_{\perp},\mathcal{M}_{\parallel}$), the Venn diagram in Fig.~\ref{fig2}(b) becomes symmetric between the left and right sides, where the dual relation, $m\leftrightarrow m'$ and $\bar{n}\leftrightarrow\bar{n}'$ in M-PGs, hold.
On the other hand, as the combinations of $\bm{M}$ and $\bm{T}$ appear differently in $\bm{G}$ and $\bm{Q}$, the Venn diagram in Fig.~\ref{fig2}(a) becomes asymmetric.

\subsection{Cross correlations}

\begin{table}[t!]
\caption{
\label{tbl2}
Symmetry correspondences among four-type of multipoles ($l=0,1$), conjugated fields, and responses or ordered states.
}
\scriptsize
\begin{center}
\begin{tabular}{ccc}
\br
MP & Conjugate field & Response/Order \\
\br
$Q_{0}$ & $\bm{\nabla}\cdot\bm{E}$ & $\delta T$ (temperature difference), charge \\
$M_{0}$ & $\bm{E}\cdot\bm{B}$ & monopole-type AFM \\
$G_{0}$ & $\bm{E}\cdot(\bm{\nabla}\times\bm{E})$, $\bm{B}\cdot(\bm{\nabla}\times\bm{B})$ (optical chirality; Zilch) & chirality \\
$T_{0}$ & $\bm{E}\cdot(\bm{\nabla}\times\bm{B})$, $\bm{B}\cdot(\bm{\nabla}\times\bm{E})$ & MT-monopole-type AFM \\
\mr
$\bm{Q}$ & $\bm{E}$ (electric field), $-\bm{\nabla}T$ (temperature gradient) & $\bm{P}$ (electric polarization), $\bm{u}$ (displacement), ferroelectricity \\
$\bm{M}$ & $\bm{B}$ (magnetic field) & $\bm{M}$ (magnetization), FM \\
$\bm{G}$ & $\bm{\nabla}\times\bm{E}$, $\partial\bm{B}/\partial t$ & $\bm{\nabla}\times\bm{u}$ (static rotational distortion), ferroaxial  \\
$\bm{T}$ & $\bm{\nabla}\times\bm{B}$, $\partial\bm{E}/\partial t$ & $\bm{J}$ (electric current), $\bm{J}_{\rm T}$ (thermal current), vortex-type AFM \\
\br
\end{tabular}
\end{center}
\end{table}

As mentioned above, the toroidal multipoles can coexist with other multipoles according to the symmetry correspondence.
Even in the high-symmetry PGs such as nonmagnetic grey groups where the equilibrium value of magnetic-type multipoles should vanish, the electric-type multipole can couple with magnetic-type multipoles as a fluctuation.
For example, there exists $G_{0}$ in $321'$, which can couple with $\bm{T}\cdot\bm{M}$ as it has the same symmetry properties of $G_{0}$.
It also couples with $\bm{Q}\cdot\bm{G}$ as well.
In this case, the Landau free energy can be written as
\begin{equation}
F=\frac{\chi^{-1}_{\bm{T}}}{2}\bm{T}^{2}+\frac{\chi^{-1}_{\bm{M}}}{2}\bm{M}^{2}
-\bm{M}\cdot\bm{H}
-\bm{T}\cdot(\bm{\nabla}\times\bm{H})
-\alpha G_{0}(\bm{T}\cdot\bm{M})+\cdots,
\end{equation}
where $\chi_{A}$ is the susceptibility for $A=\bm{M},\bm{T}$ and $\alpha$ is the trilinear coupling constant.
The third and fourth terms represent the interaction between $A$ and its conjugate field where $\bm{H}$ is an external magnetic field.
The equilibrium values for $\bm{M}$ and $\bm{T}$ in the presence of $G_{0}$ can be obtained by minimizing $F$ with respect to them as
\begin{equation}
\left(\begin{array}{c} \bm{M} \\ \bm{T} \end{array}\right)
=\frac{\chi_{\bm{M}}\chi_{\bm{T}}}{1-\chi_{\bm{M}}\chi_{\bm{T}}(\alpha G_{0})^{2}}
\left(\begin{array}{cc} \chi_{\bm{T}}^{-1} & \alpha G_{0} \\ \alpha G_{0} & \chi^{-1}_{\bm{M}} \end{array}\right)
\left(\begin{array}{c} \bm{H} \\ \bm{\nabla}\times\bm{H} \end{array}\right).
\end{equation}
From this expression, by applying the electric current, which leads to finite $\bm{\nabla}\times\bm{H}$ along the applied current direction, the magnetic dipole $\bm{M}$ is induced linearly through the trilinear coupling.
This is nothing but the Edelstein effect, i.e., linear electric-current induced magnetization~\cite{edelstein1990spin, yoda2015current, furukawa2017observation, yoda2018orbital, Furukawa_PhysRevResearch.3.023111}.
It should be noted however that since the situation in the presence of steady current is not an equilibrium state, this type of cross correlation should be treated by the Kubo formula or Boltzmann transport equation, in a rigorous sense~\cite{Sodemann_PhysRevLett.115.216806, Gao_PhysRevLett.124.077401, Watanabe_PhysRevResearch.2.043081, Watanabe_PhysRevX.11.011001, Oiwa_doi:10.7566/JPSJ.91.014701}.

Here, there are a few comments.
In the Landau theory, the free energy is constructed under the guidance of symmetry in terms of macroscopic variables irrespective of their microscopic origin.
Thus, the argument holds for various microscopic origins of macroscopic variables arising either from cluster-type distributions of classical dipoles or pure quantum-mechanical degrees of freedom.
It is often considered the cluster-type toroidal orders where the primary variables are ordinary dipoles.
In this case, without using the concept of toroidal moment, the coupling term to a toroidal moment, such as $\bm{T}\cdot(\bm{\nabla}\times\bm{H})$ is often dropped.
Moreover, a quantum-mechanical type of toroidal order has a finite thermal average of the corresponding toroidal operator itself as discussed in the previous section without any accompanying dipole moments.

In the above discussion, the trilinear term determines possible linear cross correlations.
As was already stated, $G_{0}$ representing chiral state can couple with $T_{0}M_{0}$, $\bm{Q}\cdot\bm{G}$, and $\bm{T}\cdot\bm{M}$.
At the same time, the corresponding composite fields of $\bm{Q}\cdot\bm{G}$ and $\bm{T}\cdot\bm{M}$, i.e., $\bm{E}\cdot(\bm{\nabla}\times\bm{E})$ and $\bm{B}\cdot(\bm{\nabla}\times\bm{B})$, whose sum is known as the optical chirality (Zilch)~\cite{lipkin1964existence, kibble1965conservation, proskurin2017optical} and is activated by circularly polarized light, represent the conjugate field of $G_{0}$.
These couplings imply various cross correlations under chiral state, such as the electric-field induced rotational distortion ($\bm{E}\to\bm{G}$)~\cite{Oiwa_PhysRevLett.129.116401, hayami2023chiral}, the current-induced Magnetization (Edelstein effect) ($\bm{J}\to\bm{M}$)~\cite{edelstein1990spin, yoda2015current, furukawa2017observation, yoda2018orbital, Furukawa_PhysRevResearch.3.023111}, and the magnetic-field induced vortex-like spin alignment ($\bm{B}\to\bm{T}$).

Similarly, $\bm{G}$ representing ferroaxial state can couple with $G_{0}\bm{Q}$, $T_{0}\bm{M}$, $M_{0}\bm{T}$, and $\bm{X}_{1}\times\bm{X}_{2}$.
These couplings under ferroaxial state lead to the electric-field induced chirality (electro-gyration) ($\bm{E}\to G_{0}$)~\cite{hayashida2020visualization, Hayashida_PhysRevMaterials.5.124409}, the magnetic-field induced MT monopole ($\bm{B}\to T_{0}$), and the transverse responses by fields ($\bm{X}\to \bm{G}\times\bm{X}$)~\cite{Hayami_doi:10.7566/JPSJ.91.113702, Nasu_PhysRevB.105.245125}.
The last response will be discussed later.

The MT-monopole $T_{0}$ order can be realized in antiferromagnets (AFMs) with proper structures as shown later~\cite{Hayami_PhysRevB.108.L140409}.
In this case, $T_{0}$ can couple with $G_{0}M_{0}$, $\bm{G}\cdot\bm{M}$, and $\bm{Q}\cdot\bm{T}$.
Therefore, we expect the time-reversal switching cross correlations, such as the static electromagnetic-field induced chirality ($\bm{E}\cdot\bm{B}\to G_{0}$), the magnetic-field induced static rotational distortion ($\bm{B}\to\bm{G}$), and electric-field induced vortex-like AFM ($\bm{E}\to \bm{T}$).
The MT-dipole $\bm{T}$ orders in some AFMs and the associated cross correlations have already been discussed extensively~\cite{Spaldin_0953-8984-20-43-434203, kopaev2009toroidal, Hayami_PhysRevB.90.024432, Fiebig0022-3727-38-8-R01, khomskii2009classifying, Gao_PhysRevB.98.060402, Shitade_PhysRevB.98.020407, Yatsushiro_PhysRevB.105.155157}.
$\bm{T}$ couples with $\bm{G}\cdot\bm{T}$, $\bm{Q}\cdot\bm{M}$ and so on.
Among them, the magneto-electric effect ($\bm{E}\to\bm{M}$ or $\bm{B}\to\bm{Q}$) is the representative cross correlation.

\begin{figure}[tb]
\centering
\includegraphics[width=9cm]{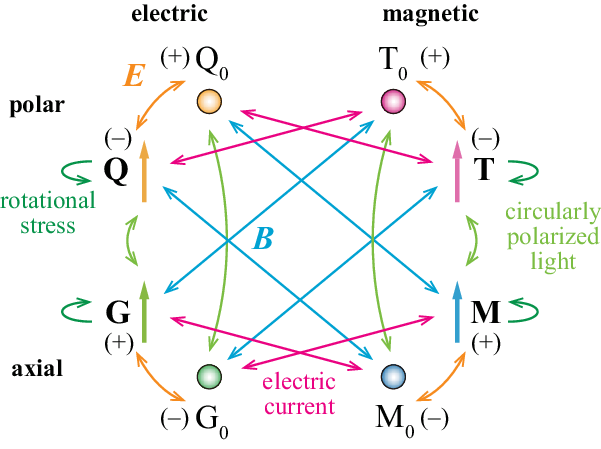}
\caption{\label{fig3}
Possible conversion among monopoles and dipoles by external fields: electric field ($\bm{E}$), magnetic field ($\bm{B}$), electric current, rotational stress, and circularly polarized light.
The sign in the parenthesis indicates the spatial parity of multipole.
}
\end{figure}

The symmetry correspondence among multipoles ($l=0,1$), conjugate fields, and responses or ordered states, and conversion among multipoles by external fields are summarized in Table~\ref{tbl2} and Fig.~\ref{fig3}.
When we consider such analysis for multipoles with $l>1$, e.g., second-rank stress field ($Q_{2m}$), spin current $J_{i}M_{j}$, etc., we expect a further variety of cross correlations with toroidal states in a unified way.

\section{Examples}\label{sec4}

\subsection{Chirality as electric-toroidal monopole $G_{0}$}

As was already mentioned, from a symmetry point of view the definition of chirality matches an ET monopole, $G_{0}$~\cite{Hayami_PhysRevB.98.165110, kusunose2020complete, Oiwa_PhysRevLett.129.116401, kishine2022definition}.
For example, the typical chiral crystal is elemental tellurium (Te), which belongs to the trigonal space group $P3_{1}21$ (\#152,$D_{3}^{4}$) or $P3_{2}21$ (\#154,$D_{3}^{6}$) as a pair of right-handed and left-handed enantiomers.
In the unit cell, three sublattices constitute the threefold helical chain, representing three-dimensional structural chirality.

\begin{figure}[tb]
\centering
\includegraphics[width=15cm]{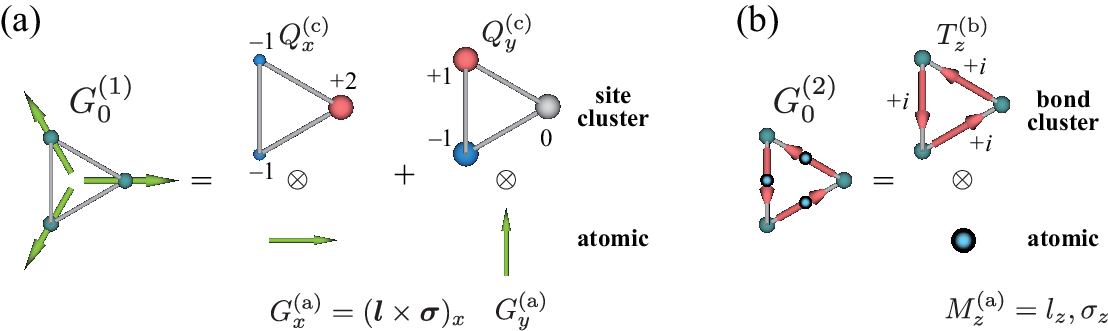}
\caption{\label{figte}
Schematic representations of chirality $G_{0}$, (a) site-cluster type $G_{0}^{(1)}$ and (b) bond-cluster (itinerant) type $G_{0}^{(2)}$~\cite{Oiwa_PhysRevLett.129.116401}.
}
\end{figure}

The electronic states around H point in the hexagonal Brillouin zone near the Fermi level consist of $p$ orbitals, and they are analyzed by the symmetry-adapted multipole basis~\cite{Oiwa_PhysRevLett.129.116401}.
The results show that three types of $G_{0}$ give dominant contributions to the tight-binding model of Te,
\begin{equation}
G_{0}^{(1)}=\frac{1}{\sqrt{2}}(Q_{x}^{\rm (c)}G_{x}^{\rm (a)}+Q_{y}^{\rm (c)}G_{y}^{\rm (a)}),
\quad
G_{0}^{(2)}=T_{z}^{\rm (b)}M_{z}^{\rm (a)},
\quad
G_{0}^{(3)}=\frac{1}{\sqrt{2}}(T_{x}^{\rm (b)}M_{x}^{\rm (a)}+T_{y}^{\rm (b)}M_{y}^{\rm (a)}),
\end{equation}
where the superscripts, c, b, and a, represent cluster, bond, and atomic bases, respectively, and schematically shown in Fig.~\ref{figte}.
Here, $\bm{M}^{\rm (a)}$ mainly arises from spin degrees of freedom.
$G_{0}^{(1)}$ represents divergent alignment of $\bm{G}^{(a)}\propto \bm{l}\times\bm{\sigma}$, while $G_{0}^{(2)}$ and $G_{0}^{(3)}$ describe the spin-dependent imaginary hopping processes along and perpendicular to the helical chain, respectively.
The latter $G_{0}^{(2)}$ and $G_{0}^{(3)}$ are the microscopic origin of Edelstein effect observed by the spectrum shift of $^{125}$Te in NMR measurement~\cite{furukawa2017observation, Furukawa_PhysRevResearch.3.023111}.
On the other hand, $G_{0}^{(1)}$ can contribute another cross correlation, such as the electro-rotation (electric-field induced rotational distortion) effect ($\bm{E}\to\bm{G}$) as discussed in the previous section, since the expression consists of $\bm{Q}$ and $\bm{G}$~\cite{Oiwa_PhysRevLett.129.116401}.
Once the existence of such microscopic couplings was confirmed by the experiments, one could achieve absolute enantioselection by using composite conjugate fields to access $G_{0}$ through such couplings.
Note that the ET quadrupole $G_{u}$ also belongs to the identity representation of $D_{3}$, which gives the monoaxial anisotropy.

There is another attempt to quantify the chirality of molecule by using the ET monopole.
For such purpose, a twisted methane is analyzed as a prototype of the method, and the results indeed show that the ET monopole becomes a quantitative indicator for chirality~\cite{inda2024quantification}.
In the twisted methane, it is clarified that the handedness of chirality corresponds to the sign of the expectation value of the electric toroidal monopole, and that the most important ingredient is the modulation of the spin-dependent imaginary hopping between the hydrogen atoms, while the relativistic spin-orbit coupling within the carbon atom is irrelevant for chirality.
The method can be straightforwardly applied to more realistic chiral molecules, which would unveil the nature of chirality quantitatively at the quantum-mechanical level.

\subsection{Ferroaxial order ($\bm{G}$), and electro-gyration effect and spin current generation}

\begin{figure}[tb]
\centering
\includegraphics[width=14cm]{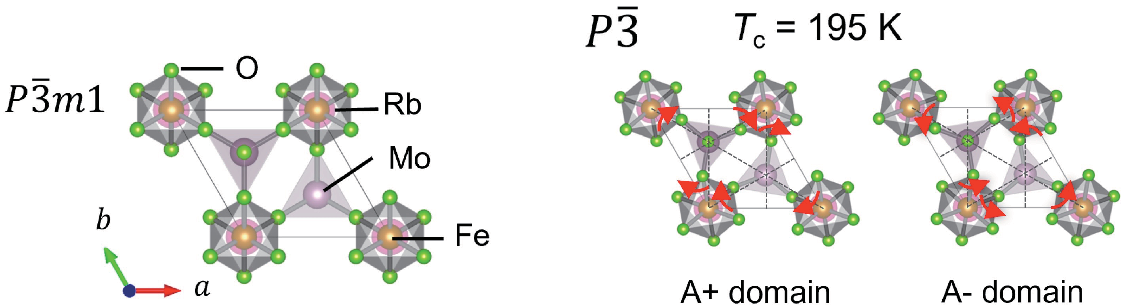}
\caption{\label{fig4}
Crystal structure of RbFe(MoO$_{4}$)$_{2}$ in (left) normal and (right) ferroaxial phases.
The twin domains appear in the ordered phase depending on the rotated direction of octahedra and tetrahedra units.
Reprinted and modified figures with permission from~\cite{Hayashida_PhysRevMaterials.5.124409}, \copyright 2021 American Physical Society.
}
\end{figure}

\begin{figure}[tb]
\centering
\includegraphics[width=11cm]{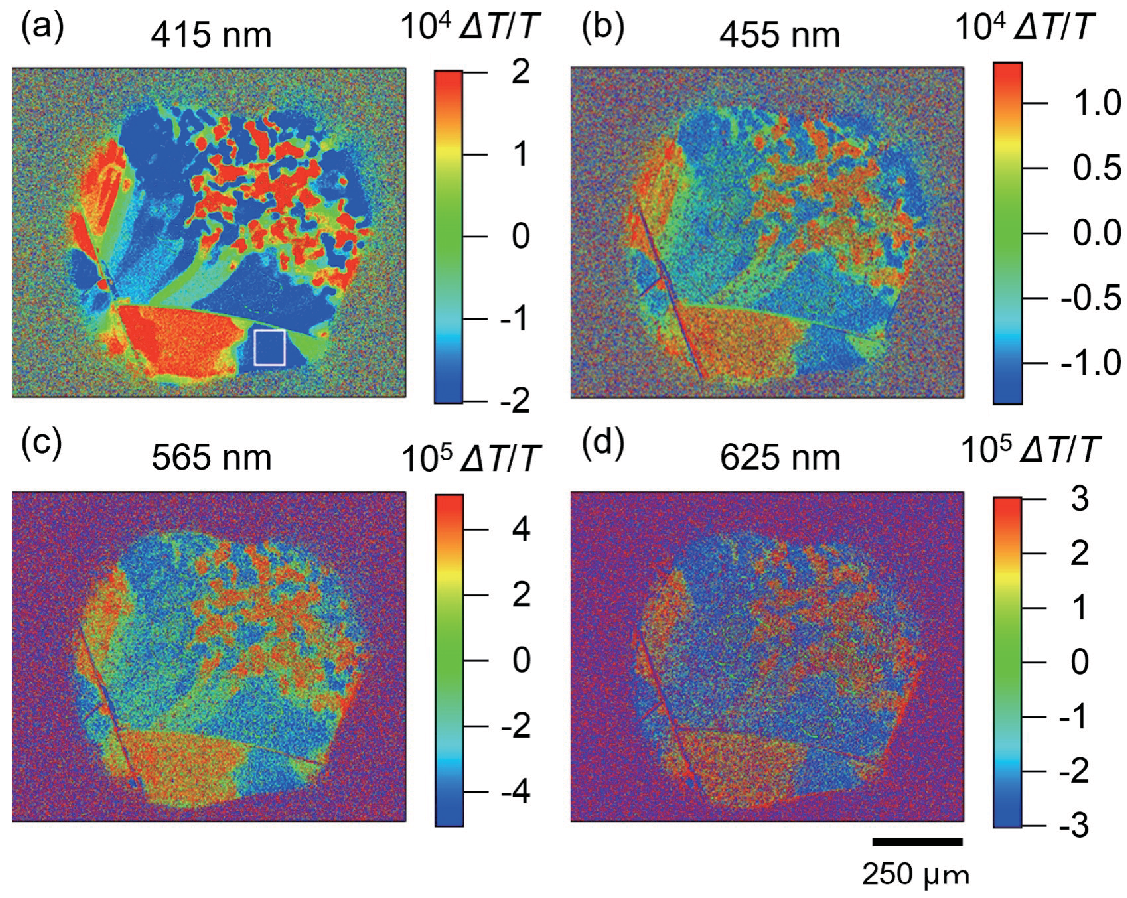}
\caption{\label{fig5}
Imaging of the ferroaxial domain structure of RbFe(MoO$_{4}$)$_{2}$ by the optical rotation induced by electro-gyration effect, where $\Delta T/T$ is the relative rotation angle of the light polarization. Reprinted figures with permission from~\cite{Hayashida_PhysRevMaterials.5.124409}, \copyright 2021 American Physical Society.
}
\end{figure}

The ferroaxial order is described by the uniform alignment of $\bm{G}$.
RbFe(MoO$_{4}$)$_{2}$ is one of the typical crystals showing the ferroaxial order.
The crystal structure of this compound is shown in the left panel of Fig.~\ref{fig4}, where FeO$_{6}$ octahedra sharing vertices with MoO$_{4}$ tetrahedra is stacked along $c$ axis.
It exhibits the structural phase transition at $T_{c}\sim 195$ K, in which octahedra and tetrahedra rotate in opposite direction around $c$ axis as shown in the right panel of Fig.~\ref{fig4}, and the point group changes from $\bar{3}m$ ($D_{\rm 3d}$) to $\bar{3}$ ($C_{\rm 3i}$) with losing vertical mirror operations.
This structural transition has been detected by the high-sensitive rotational second-harmonic generation method, in which the electric-quadrupole transition is analyzed~\cite{jin2020observation}.

In the ET dipole order $\bm{G}$, the electric field can induce chirality $G_{0}$ (electro-gyration effect: $\bm{E}\to G_{0}$), and the enantioselection, i.e., the sign of $G_{0}$, can be realized by the direction of the electric field as long as the single domain is assumed.
In reality, however, there are twin domains, and the electro-gyration effect is utilized for domain detection as shown in Fig.~\ref{fig5} for RbFe(MoO$_{4}$)$_{2}$, because the optical rotation angle of the linear polarized light depends on the domain of $\bm{G}$ with the fixed applied electric field~\cite{Hayashida_PhysRevMaterials.5.124409}.
The visualization of the ferroaxial domain is particularly important as the achievement of the enantioselection strongly depends on how the domain is distributed in the sample.

A similar method is also applied to NiTiO$_{3}$, which has elucidated that the domain size depends on the cooling rate around $T_{c}$~\cite{hayashida2020visualization, Hayashida_PhysRevMaterials.5.124409, fang2023ferrorotational}.
Moreover, state-of-the-art three-dimensional imaging by circularly polarized second harmonic generation microscopy has also been developed to observe the ET dipole in real space~\cite{yokota2022three}.
These experimental techniques further stimulate the detection of $\bm{G}$ in other candidates, such as Co$_{3}$Nb$_{2}$O$_{8}$~\cite{Johnson_PhysRevLett.107.137205}, CaMn$_{7}$O$_{12}$~\cite{Johnson_PhysRevLett.108.067201}, Ca$_{5}$Ir$_{3}$O$_{12}$~\cite{Hasegawa_doi:10.7566/JPSJ.89.054602, Hanate_doi:10.7566/JPSJ.89.053601, hanate2021first, hanate2023space}, BaCoSiO$_{4}$~\cite{Xu_PhysRevB.105.184407}, K$_{2}$Zr(PO$_{4}$)$_{2}$~\cite{yamagishi2023ferroaxial}, Na$_{2}$Hf(BO$_{3}$)$_{2}$~\cite{nagai2023chemicalSwitching}, and the typical candidates are Ca(BH$_{4}$)$_{2}$~\cite{filinchuk2009crystal}, PdS~\cite{brese1985reinvestigation}, TaPO$_{5}$~\cite{longo1971tetragonal}, VPO$_{5}$~\cite{gautier2013dft}, and NbTe$_{2}$~\cite{bhatt2004x} belonging to $C_{\rm 4h}$ point group.

\begin{figure}[tb]
\centering
\includegraphics[width=14cm]{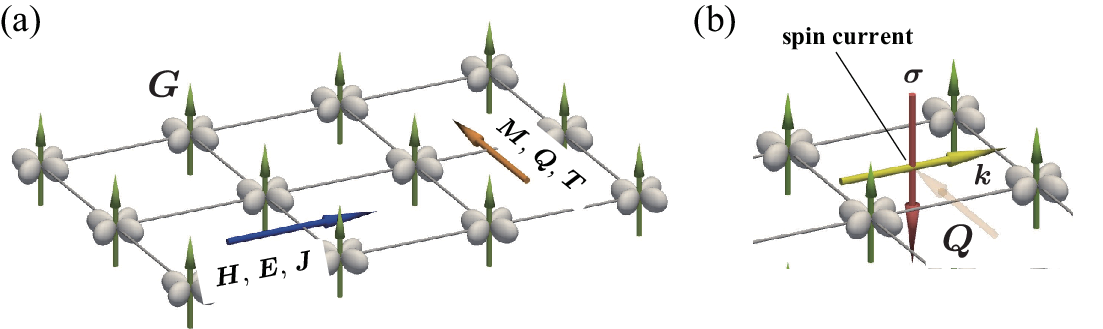}
\caption{\label{fig_rot}
(a) Ferroic $\bm{G}$ order (green arrows) plays a role of a ``nanorotator'', where the inputs of $\bm{H}$, $\bm{E}$, and $\bm{J}$ can be converted to the transverse output of $\bm{M}$, $\bm{Q}$, and $\bm{T}$, respectively.
The canted orbitals schematically show the broken vertical mirror symmetry, which accommodates the $\bm{G}$ order.
(b) The electric dipole $\bm{Q}$ in the momentum space is the same symmetry as the spin current perpendicular to $\bm{Q}$, and the $\bm{G}$ order provides a playground for the longitudinal spin current.
This figure is taken and modified from~\cite{Hayami_doi:10.7566/JPSJ.91.113702}.
}
\end{figure}

In addition to the above conversion from dipole to pseudoscalar, the conversion among dipoles can also be expected in $\bm{G}$ order owing to the coupling between $\bm{G}$ and $\bm{X}_{1}\times\bm{X}_{2}$~\cite{Hayami_doi:10.7566/JPSJ.91.113702}.
This cross correlation is characterized by the conversion where the input and output vectors are orthogonal with each other, and hence $\bm{G}$ order plays a role of a nanorotator as shown in Fig.~\ref{fig_rot}(a).
Therefore, the inputs of $\bm{H}$, $\bm{E}$, and $\bm{J}$ can be converted to the transverse output of $\bm{M}$, $\bm{Q}$, and $\bm{T}$, respectively.
In particular, as the electric dipole $\bm{Q}$ in the momentum space is expressed as $\bm{Q}\propto \bm{k}\times\bm{\sigma}$, the transverse output is equivalent to the longitudinal spin current along the applied electric field as shown in Fig.~\ref{fig_rot}(b).
The experimental confirmation of such cross correlation is highly desired.

\subsection{Magneto-electric gyration under magnetic-toroidal monopole order ($T_{0}$)}

\begin{table}[t!]
\caption{
\label{tbl3}
Classification for magnetic-toroidal monopole order ($T_{0}$) in Type I magnetic point groups and candidate materials~\cite{Hayami_PhysRevB.108.L140409}.
Other multipoles indicated by open circles are accompanied with $T_{0}$ order.
}
\small
\begin{center}
\begin{tabular}{cccccc}
\br
PG & $T_{0}$ & $M_{z}$ & $T_{z}$ & $M_{0}$ & Materials \\
\br
$O_{\rm h}$, $T_{\rm d}$, $T_{\rm h}$, $D_{\rm 2h}$, $D_{\rm 3h}$, $D_{\rm 4h}$, $D_{\rm 6h}$, $D_{\rm 2d}$, $D_{\rm 3d}$ & $\circ$ & & & & KMnF$_{3}$, Ca$_{2}$RuO$_{4}$ \\
$C_{\rm 2h}$, $C_{\rm 3h}$, $C_{\rm 4h}$, $C_{\rm 6h}$, $S_{4}$, $C_{\rm 3i}$, $C_{\rm i}$ & $\circ$ & $\circ$ & & & MnV$_{2}$O$_{4}$, Mn$_{3}$As$_{2}$ \\
$C_{\rm 2v}$, $C_{\rm 3v}$, $C_{\rm 4v}$, $C_{\rm 6v}$ & $\circ$ & & $\circ$ & & YMnO$_{3}$, Er$_{2}$Cu$_{2}$O$_{5}$ \\
$O$, $T$, $D_{2}$, $D_{3}$, $D_{4}$, $D_{6}$ & $\circ$ & & & $\circ$ & Ho$_{2}$Ge$_{2}$O$_{7}$, Mn$_{3}$IrGe \\
$C_{\rm s}$ & $\circ$ & $\circ$ & $\circ$ & & Mn$_{4}$Nb$_{2}$O$_{9}$ \\
$C_{1}$, $C_{2}$, $C_{3}$, $C_{4}$, $C_{6}$ & $\circ$ & $\circ$ & $\circ$ & $\circ$ & ScMnO$_{3}$, Mn$_{2}$FeMoO$_{6}$ \\
\br
\end{tabular}
\end{center}
\end{table}

\begin{figure}[tb]
\centering
\includegraphics[width=10cm]{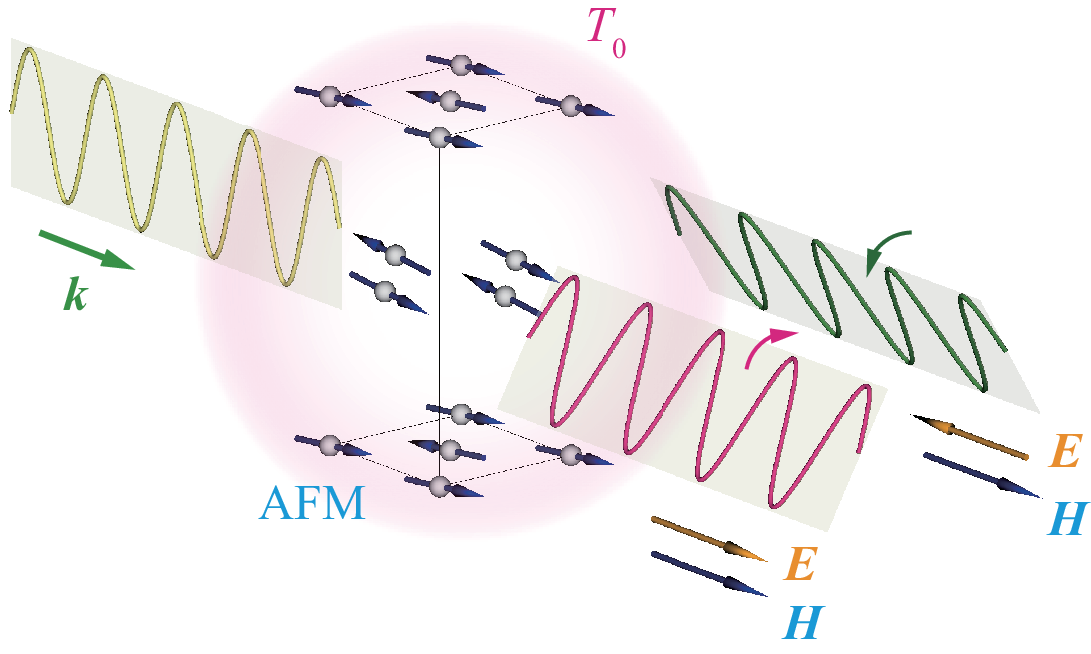}
\caption{\label{fig_cro}
Magnetic structure in Ca$_{2}$RuO$_{4}$ accompanying $T_{0}$, and expected optical rotation measurement in the applied electric and magnetic fields, $\bm{E}$ and $\bm{H}$.
The rotation angle depends on the parallel or anti-parallel fields of $\bm{E}$ and $\bm{H}$.
Reprinted figures with permission from~\cite{Hayami_PhysRevB.108.L140409}, \copyright 2023 American Physical Society.
}
\end{figure}

\begin{table}[t!]
\caption{
\label{tbl_cro}
Magnetic moments at four Ru sites in Ca$_{2}$RuO$_{4}$~\cite{Porter_PhysRevB.98.125142}. 
$\bm{T}^{\rm (c)}_u$, $\bm{T}^{\rm (c)}_v$, and $\bm{M}^{\rm (c)}_{xyz}$ represent the cluster $3z^2-r^2$-type and $x^2-y^2$-type MT quadrupoles, and $xyz$-type magnetic octupole, respectively, which are all belonging to the identity representation.
}
\small
\begin{center}
\begin{tabular}{ccccccc}
\br
Position &  $\bm{M}$ &  $\bm{T}^{\rm (c)}_u$&  $\bm{T}^{\rm (c)}_v$ &  $\bm{M}^{\rm (c)}_{xyz}$ \\
\br
$(0,0,0)$ & $(0,+1.0,+0.1)$ & $\frac{1}{2\sqrt{2}}(1,-1,0)$ & $\frac{1}{2\sqrt{6}}(1,1,-2)$  & $\frac{1}{2\sqrt{3}}(1,1,1)$ \\
$(1/2,0,1/2)$ & $(0,-1.0,-0.1)$ & $\frac{1}{2\sqrt{2}}(-1,1,0)$ & $\frac{1}{2\sqrt{6}}(-1,-1,-2)$  & $\frac{1}{2\sqrt{3}}(-1,-1,1)$ \\
$(0,1/2,1/2)$ & $(0,+1.0,-0.1)$ & $\frac{1}{2\sqrt{2}}(-1,-1,0)$ & $\frac{1}{2\sqrt{6}}(-1,1,2)$ & $\frac{1}{2\sqrt{3}}(-1,1,-1)$ \\
$(1/2,1/2,0)$ & $(0,-1.0,-0.1)$ &$\frac{1}{2\sqrt{2}}(1,1,0)$ & $\frac{1}{2\sqrt{6}}(1,-1,2)$ & $\frac{1}{2\sqrt{3}}(1,-1,-1)$ \\
\br
\end{tabular}
\end{center}
\end{table}

As discussed in the previous section, a divergent alignment of the vortex-like AFM represents the cluster-type MT monopole, $T_{0}$~\cite{Hayami_PhysRevB.108.L140409}.
Indeed, AFMs belonging to the same paramagnetic point group only without $\mathcal{T}$ operation, i.e., Type I magnetic point group, can exhibit the MT monopole order, and they accompany additional multipoles in low-symmetry point groups.
The classification of active multipoles and the candidate AFMs are summarized in Table~\ref{tbl3}.

Among them, Ca$_{2}$RuO$_{4}$ for example shows AFM order ($mmm1'\to mmm$, $D_{\rm 2h}$) as shown in Fig.~\ref{fig_cro}, and the ordered moment at each position is given by $\bm{M}$ in Table~\ref{tbl_cro}.
There are nine cluster-type magnetic multipoles, $\bm{T}_{u}^{\rm (c)}$, $\bm{T}_{v}^{\rm (c)}$, and $\bm{M}_{xyz}^{\rm (c)}$, belonging to the identity representation, and the observed magnetic structure can be expressed as the linear combination as
\begin{equation}
\bm{M}=-1.414\bm{T}_{u}^{\rm (c)}+0.653\bm{T}_{v}^{\rm (c)}+1.270\bm{M}_{xyz}^{\rm (c)}.
\end{equation}
Therefore, we expect various cross correlations in the presence of $T_{0}$  as discussed previously.
For example, a possible experimental setup is shown in Fig.~\ref{fig_cro}, which is expected to show the optical rotation in the presence of static electric and magnetic fields, and the rotation angle depends on the relative directions between electric and magnetic fields.
The direction of incident light can be chosen arbitrarily since the order parameter has a scalar nature.
In this experiment, it would be also important to know how the domain of $T_{0}$ is distributed in the sample.
This type of experiment also provides a way of absolute enantioselection by static electric and magnetic fields as long as the single domain is assumed.

\section{Summary}\label{sec5}

In summary, we have introduced the concept of toroidal multipoles both from classical and quantum viewpoint.
By using the concept, we have shown the correspondence between active multipoles and crystallographic point groups in non-magnetic materials, and magnetic point groups in magnetic materials, with focusing on monopoles and dipoles.
The electric-toroidal monopole is a practical entity of chirality, while the electric-toroidal dipole is a microscopic variable of ferroaxial order.
The magnetic-toroidal monopole and dipole are often realized in certain type of antiferromagnets.
Once the relevant symmetry-adapted multipoles are identified in materials, we can expect various cross correlations in a unified way at the quantum-mechanical level.
The proper understanding of these toroidal moments would stimulate further investigation of mutual conversions among peculiar states of matter.

\section*{Acknowledgement}

This paper is based on fruitful collaborations with Yuki Yanagi, Megumi Yatsushiro, Rikuto Oiwa, Jun-ichiro Kishine, and Hiroshi H. Yamamoto.
We would also like to thank Makoto Naka, Michi-to Suzuki, Yukitoshi Motome, Hitoshi Seo, Takuya Nomoto, Ryotaro Arita, Hiroshi Amitsuka, Tatsuya Yanagisawa, Takahisa Arima, Hiraku Saito, for helpful discussions.
This research was supported by JSPS KAKENHI Grants Numbers JP21H01031, JP21H01037, JP22H04468, JP22H00101, JP22H01183, JP23K03288, JP23H04869, JP23H00091 and by JST PRESTO (JP- MJPR20L8) and JST CREST (JPMJCR23O4), and the grants of Special Project (IMS program 23IMS1101), and OML Project (NINS program No. OML012301) by the National Institutes of Natural Sciences.

\vspace{5mm}

\bibliographystyle{iopart-num}
\bibliography{ref}

\end{document}